\documentclass[twocolumn]{aastex631}
\usepackage{color}
\usepackage{threeparttable}

\definecolor{mygreen}{rgb}{0.19,0.55,0.11}

\shorttitle{Formation of PTF~J0533+0209}
\shortauthors{Chen et al.}
\graphicspath{{./}{figures/}}

\begin{document}

\title{Formation of the double white dwarf binary PTF~J0533+0209 through stable mass transfer?}
\correspondingauthor{Hai-Liang Chen}
\email{chenhl@ynao.ac.cn}

\author{Hai-Liang Chen}
\affiliation{Yunnan Observatories, Chinese Academy of Sciences (CAS), Kunming 650216, P.R. China}
\affiliation{Department of Physics and Astronomy, Aarhus University, Ny Munkegade 120, 8000~Aarhus~C, Denmark}

\author[0000-0002-3865-7265]{Thomas M. Tauris}
\affiliation{Department of Physics and Astronomy, Aarhus University, Ny Munkegade 120, 8000~Aarhus~C, Denmark}

\author{Xuefei Chen}
\affiliation{Yunnan Observatories, Chinese Academy of Sciences (CAS), Kunming 650216, P.R. China}
\affiliation{University of the Chinese Academy of Sciences, Yuquan Road 19, Shijingshan Block, 100049, Beijing, China}

\author[0000-0001-9204-7778]{Zhanwen Han}
\affiliation{Yunnan Observatories, Chinese Academy of Sciences (CAS), Kunming 650216, P.R. China}
\affiliation{University of the Chinese Academy of Sciences, Yuquan Road 19, Shijingshan Block, 100049, Beijing, China}

\begin{abstract}
Double white dwarf (DWD) binaries are important for studies of common-envelope (CE) evolution, Type~Ia supernova progenitors and Galactic sources of low-frequency gravitational waves (GWs). PTF~J0533+0209 is a DWD system with a short orbital period of $P_{\rm orb} \sim 20\;$min and thus a so-called LISA verification source. The formation of this system and other DWDs is still under debate. In this paper, we discuss the possible formation scenarios of this binary and argue that it is not likely to have formed through CE evolution. Applying a new magnetic braking prescription, we use the \texttt{MESA} code to model the formation of this system through stable mass transfer. 
We find a model which can well reproduce the observed WD masses and orbital period but not the effective temperature and hydrogen abundance of the low-mass He~WD component. 
We discuss the possibility of using H flashes to 
mitigate this discrepancy.
Finally, we discuss the future evolution of this system into a AM~CVn binary such as those that will be detected by space-borne GW observatories like LISA, TianQin and Taiji. 
\end{abstract}


\section{Introduction}
\label{sec:intro}

Recently, a number of double white dwarfs (DWDs) with very short orbital periods have been discovered, such as ZTF~J1539+5027 ($\sim 7\;$min orbital period, \citealt{bcfk+19} and PTF~J0533+0209 ($\sim 20\;$min orbital period, \citealt{bfpv+19}). In these DWDs, the primary WD is a CO~WD and the companion star is an extremely low-mass (ELM) He~WD with a mass smaller than $\sim 0.2\;M_{\odot}$. Because of their short orbital periods, they are important Galactic gravitational wave (GW) sources that will be detected by space-borne observatories, such as LISA \citep{aabb+17}, TianQin \citep{lcdg+16}, Taiji \citep{rgcz20}. However, the formation of these systems is not well understood. 

The formation of DWDs has been widely investigated in the past \citep{it86,lp88,ty94,ylts94,hpe95,ity97,han98,sa18,lcch19}.
It is generally accepted that there are two possible formation scenarios for
DWDs (see also Fig. 2 in \citealt{lcch19}). In the first formation scenario, a binary system consisting of a CO~WD and a low-mass donor star undergoes stable mass transfer via Roche-lobe overflow (RLO) and evolves into a binary system consisting of a CO~WD and a He~WD. The systems produced from this scenario will follow the He~WD mass--orbital period relation \citep[e.g.][]{ts99}. 
In the second scenario, a binary system with a CO~WD and a red-giant companion star is subject to unstable mass transfer and enters a common-envelope (CE) phase. If the system successfully ejects the CE, then it will evolve into a CO~WD+He~WD binary. Hereafter, we refer to these two scenarios as the stable RLO scenario and the CE scenario, respectively. 

\citet{sa18} have studied the formation of ELM~WDs from binary evolution using a mass range of the progenitors of the ELM~WDs between $1.0-1.50\;M_{\odot}$. They found that ELM~WDs can be produced from non-conservative mass transfer and are not likely to be formed via CE evolution. 

\citet{lcch19} have comprehensively studied the formation of ELM~WDs through both stable RLO and CE evolution. 
They found that the ELM WDs can form from both scenarios and argue that ELM~WDs with a mass smaller than $\sim 0.22\;M_{\odot}$ are formed from the stable RLO scenario, whereas systems with ELM~WDs $>0.22\;M_\odot$ formed via the CE scenario.

Applying so-called convection and rotation-boosted magnetic braking, \citet{sk21} studied the formation of ELM~WDs with NS companions and found that ELM~WD masses of $0.15-0.27\;M_{\odot}$ can be 
produced via stable RLO from low-mass X-ray binaries (LMXBs) with initial orbital periods of $1-25\;{\rm d}$.

For the very tight-orbit DWD system with an ELM~He~WD component above a certain mass limit the origin is clear. As an example, one may consider ZTF~J1539+5027 with an orbital period of $\sim 7\;$min and an ELM~He~WD mass of $\sim 0.21\;M_{\odot}$. If this system were produced from stable RLO, then the orbital period at the end of mass transfer would have been around $2.4\;$d according to the WD mass---orbital period relation from \citet{ts99}. It would subsequently have taken the system around $3.4\times 10^{12}\;$yr to evolve to the current orbital period of $7\;$min via GW damping. Therefore, this system cannot have been produced from stable RLO. 

Regarding the system PTF~J0533+0209, which is the focus of this paper (see Table~\ref{tab:para_com} for parameters), \citet{bfpv+19} suggested that it is formed through a CE scenario. But they did not discuss whether the CE can be ejected in order to produce such a system. The results of \citet{sa18} and \citet{lcch19} show that this system is not likely to be formed from a CE scenario. 
On the other hand, it is unclear from the binary models of \citet{lcch19,sk21} if PTF~J0533+0209 can be produced from stable RLO and subsequently decay its orbit sufficiently (once detaching from the WD mass—orbital period correlation) within a Hubble time to reproduce its present orbital period of 20.6~min.
Therefore, this system is puzzling as it is apparently also hard to explain its origin with binary evolution models with stable RLO.
Here we aim to investigate this system in more detail as it represents a class of DWD systems challenging current knowledge of binary stellar evolution.

Recently, \citet{cthc21} investigated the formation of binary ELM~WDs with radio millisecond pulsar (MSP) companions. Using an intermediate strength magnetic-braking prescription (``MB3'') suggested by \citet{vih19}, they found that the initial parameter space of LMXBs for producing such MSP+ELM~WD binaries becomes significantly larger compared to models with a standard magnetic braking, and thereby offering a potential solution to the ``fine-tuning problem'' of producing MSP+ELM~WD binaries \citep{itl14}. However, they also found that with the MB3 prescription they cannot reproduce many known wide-orbit binary MSPs. Motivated by this work, we will here study the formation of the system PTF~J0533+0209 using a magnetic braking prescriptions from \citet{vih19}.

The purpose of this paper is to discuss possible formation scenarios for PTF~J0533+0209 and present a valid binary formation model for this system. The paper is structured as follows. In Section~\ref{sec:ce}, we demonstrate that formation of PTF~J0533+0209 through CE evolution is not possible from a simple energy budget argument. 
In Section~\ref{sec:bem}, we discuss the formation of this system through stable RLO and present a viable binary evolution model for this system. We compare with observational characteristics and discuss discrepancies. In addition, we discuss the future evolution of this system as an AM~CVn system. A brief discussion and a summary of our results are given in Section~\ref{sec:dis} and Section~\ref{sec:con}, respectively. 

\begin{table}
\begin{threeparttable}
\caption{Parameters of the DWD system PTF~0533+0209. Comparison of observed data \citep{bfpv+19} and our simulated binary with $M_{\rm 2,i}=1.40\;M_\odot$, $P_{\rm orb,i}=7.00\;{\rm d}$ and $M_{\rm CO,i}=0.55\;M_\odot$.}
    \centering
    \begin{tabular}{l|l|l}
    \hline
    \hline
    \noalign{\smallskip}
    PTF~J0533+0209     & Observations & Our model  \\
    \noalign{\smallskip}
    \hline
    \noalign{\smallskip}
    Orbital period (s)            & $1234.0$                &  1234.4\\    
    CO~WD mass ($M_{\odot}$)      & $0.652\pm0.04$          &  0.669\\
    He~WD mass ($M_{\odot}$)      & $0.167\pm0.030$         &  0.155\\
    He~WD radius ($R_{\odot}$)    & $0.057\pm0.004$         &  0.060\\
    He~WD $\log g$ (cgs)          & $6.3\pm0.1^{\ast}$      &  6.1\\
    He~WD $\log (L/L_\odot)$      & $-0.33\pm0.13$          &  $-$2.24\\
    He~WD $ T_{\rm eff}$ (K)  & $20\,000\pm800$         &  6689\\
    He~WD $X_{\rm surf}$          & $4.9_{-1.0}^{+1.3}\times 10^{-4}$ & 1.000\\
    Age (Gyr)                     & ---                     & 10.660\\
    \noalign{\smallskip}
    \hline
    \hline
    \end{tabular}
    \begin{tablenotes}
    \item $^{\ast}$ Our calculated value from released data is 6.2.
    \end{tablenotes}
\end{threeparttable}    
    \label{tab:para_com}
\end{table}

\newpage
\section{Failure of a CE scenario for PTF~J0533+0209}\label{sec:ce}
In the CE scenario, a binary system consisting of a CO~WD and a (sub)red-giant star initiates unstable mass transfer and enters a CE phase. To check for the possibility of a successful CE ejection, we apply the energy budget prescription of CE evolution \citep{webb84,ls88,hpe94,hpe95}:\\
$E_{\rm bind} \equiv \alpha_{\rm CE}\,\Delta E_{\rm orb}$, or:
\begin{eqnarray}
  \int_{M_{\rm core}}^{M_{2}}\left(-\frac{G M(r)}{r}+U\right) \mathrm{d}m = \nonumber \\
\alpha_{\mathrm{CE}}\,\left(-\frac{G M_{\mathrm{core}} M_{\mathrm{CO}}}{2\, a_{\mathrm{f}}}+\frac{G(M_{\mathrm{core}}+M_{\mathrm{env}})\, M_{\mathrm{CO}}}{2\,a_{\mathrm{i}}}\right)
\label{eq:CE}
\end{eqnarray}
where the left-hand side of the equation is the binding energy, $E_{\rm bind}$ of the envelope of the giant star (with total mass $M_2=M_{\rm core}+M_{\rm env}$) before entering the CE. Here $M(r)$ is the mass within radial coordinate, $r$ of the donor star and  $U$ is the internal energy that can be used in the ejection process. The latter includes the thermal energy of a simple perfect gas, the radiation energy, as well as terms related to ionization of atoms and dissociation of molecules, as well as the Fermi energy of a degenerate electron gas \citep{hpe94,hpe95}.
The right-hand side expresses the release of orbital energy from in-spiral, $\Delta E_{\rm orb}$ and $0<\alpha _{\rm CE}<1$ is the efficiency of CE ejection (i.e. fraction of released orbital energy used to eject the CE). Here, $M_{\rm core}$, $M_{\rm env}$ and $M_{\rm CO}$ are the masses of the core, the envelope and that of the in-spiralling CO~WD, respectively, and $a_{\rm i}$ and $a_{\rm f}$ denote the binary orbital separations before and after the CE phase, respectively.
As a conservative limit, we first set $a_{\rm f}$ to be the value of current binary separation.

Following \citet{bfpv+19}, we assume that the progenitor mass of the ELM~WD is $1.20\;M_{\odot}$. Using the stellar evolution code \texttt{Modules for Experiments in Stellar Astrophysics} (\texttt{MESA}, version: 12115, \citealt{pbdh+11,pcab+13,pmsb+15,psbb+18,pssg+19}), we computed the evolution of a single star with a mass of $1.20\;M_{\odot}$ and obtain a binding energy of the envelope, $E_{\rm bind} = -1.20\times 10^{48}\;$erg and a radius, $R_{2} = 3.10\;R_{\odot}$ when the core mass is around $0.167\;M_{\odot}$ (i.e. similar to the mass of the ELM~He~WD component in the PTF~J0533+0209 system, Table~\ref{tab:para_com}). 
The initial binary separation before the CE can be computed from \citep{egg83}:
\begin{equation}
    \frac{R_{\rm rl}}{a_{\rm i}}=\frac{0.49\,q^{2/3}}{0.6\,q^{2 / 3}+\ln \left(1+q^{1 / 3}\right)},
\label{eq:rl}
\end{equation}
where $R_{\rm rl}$ is Roche-lobe radius and equal to the radius of the (sub)giant star, $q=M_{\rm donor}/M_{\rm CO}$ is the mass ratio and $M_{\rm donor}$ is the progenitor mass of the ELM WD, i.e. $M_{\rm donor} = 1.20\;M_{\odot}$. Solving Eqs.~(\ref{eq:CE}) and (\ref{eq:rl}), we find a value of $\alpha_{\rm CE} = 1.75$. 
However, this large value is not physical. Furthermore, we note that the binary separation of this system after a CE should be larger than the current orbital separation because of GW radiation since its formation until its current state, and thus the amount of released orbital energy from in-spiral would have been smaller. Therefore, the $\alpha_{\rm CE}$ value would be larger than 1.75, making the CE even more unlikely. 
It should noted that applying a larger progenitor mass of the ELM~WD would only exacerbate the problem.  
Assuming e.g. $M_{\rm donor}=1.40\;M_\odot$ increases the envelope binding energy, leading to $\alpha_{\rm CE} = 2.7$. 
Assuming a smaller mass, e.g. $M_{\rm donor} = 1.0\;M_{\odot}$ leads to longer evolution time and $\alpha_{\rm CE} = 1.14$. 

So far, the value of $\alpha_{\rm CE}$ is not well constrained (see \citealt{ijcd+13} for a review). 
\citet{dkw10} computed the detailed structure of $1-8\;M_{\odot}$ AGB stars and computed the binding energy of the envelopes of these stars. With these results, they found that it is possible to explain the population of post-CE binaries found by the SDSS with $\alpha_{\rm CE} > 0.10$. 
Using a similar approach, \citet{zsgn10} constrained the CE efficiency parameter to $0.20 < \alpha_{\rm CE} < 0.30$. 
In addition, \citet{nil15} studied the formation of DWDs from CE evolution with 3D simulations and constrained the CE efficiency parameter to $0.25 < \alpha_{\rm CE} < 0.50$. From these studies, we expect $0.1 < \alpha_{\rm CE} < 0.50$ for low-mass binaries.  
In any case, $\alpha_{\rm CE} > 1.75$ is not realistic. 
Therefore, we conclude that PTF~J0533+0209 did not form through a CE scenario and we will now focus on investigating the stable RLO scenario in the rest of the paper.

\section{stable mass transfer scenario}
\label{sec:bem}

\subsection{Method and assumptions}
Our binary evolution model is motivated by \citet{cthc21}, in which we modelled the formation of ELM~WDs in binary MSP systems via RLO in LMXBs. Here we briefly summarize the assumptions and the method we adopt in our presented calculation of PTF~J0533+0209 via RLO in a CV system. 

We use the \texttt{MESA} code to model the binary evolution. In our calculation, the CO~WD accretor is assumed to be a point mass, so we do not compute the detailed structure of this WD. The secondary star is assumed to initially be a zero-age main-sequence star with a chemical composition $X = 0.70$ (hydrogen), $Y = 0.28$ (helium) and $Z = 0.02$ (metals). We set the mixing length to be 2 times the local pressure scale height and do not consider overshooting in our calculation. The Reimer's wind prescription is used for the secondary star \citep{reim75}: 
\begin{equation}
    \dot{M}_{\rm 2,wind} = -4\times10^{-13}\;M_\odot\,{\rm yr}^{-1}\quad \eta\,\left(\frac{R_2}{R_\odot}\right) \left(\frac{L_2}{L_\odot}\right) \left(\frac{M_\odot}{M_2}\right)\;,
\end{equation}
where $\eta=1.0$ is our adopted scaling factor. $M_{2}$, $R_{2}$, $L_{2}$ are the mass, radius and luminosity of the secondary star, respectively.

In our model, we have taken different mechanisms of angular momentum loss into consideration, including angular momentum loss due to GW radiation, magnetic braking, mass loss and spin-orbital couplings.
To compute the angular momentum loss due to GW radiation, we use the formula of \citet{ll71}:
\begin{equation}
	\frac{{\rm d}J_{\rm gw}}{{\rm d}t} = -\frac{32}{5}\frac{G^{7/2}}{c^{5}}\frac{M^2_{\rm CO}M^2_2(M_{\rm CO}+M_2)^{1/2}}{a^{7/2}}\;,
\end{equation}
where G is the gravitational constant; c is the speed of light in vacuum; $M_{\rm CO}$ is the CO~WD mass; $M_{2}$ is the secondary star mass; $a$ is the orbital separation. 

Regarding the angular momentum loss due to magnetic braking, the standard Skumanich law (i.e. $\beta = \xi = \alpha = 0$ in Eq.~\ref{eq:mb_vih}) is usually applied to solar-type main sequence stars. It may be not suitable for systems with donors different than the Sun, since the stellar wind and the strength of the magnetic field can be very different which is important for the calculation of magnetic braking. By including a scaling of the strength of magnetic braking with convective turnover time and wind mass-loss rate, \citet{vih19} modelled the evolution of LMXBs and found that the observed properties of persistent X-ray binaries are in better agreement with observations using a model with $\beta = 0$, $\xi = 2$ and $\alpha =1$ (MB3), compared to models with other choices of $\beta$, $\xi$ and $\alpha$. Here, in our study of DWDs we adopt their general prescription, which can be described with the following equations:

\begin{equation}
    \frac{{\rm d}J_{\rm mb}}{{\rm d}t} = \frac{{\rm d}J_{\rm mb,Sk}}{{\rm d}t}\;\left(\frac{\omega_2}{\omega_\odot}\right)^{\beta}\left(\frac{\tau_{\rm conv}}{\tau_{\rm \odot,conv}}\right)^{\xi}\left(\frac{\dot{M}_{\rm 2,wind}}{\dot{M}_{\rm \odot,wind}}\right)^{\alpha}\;,
\label{eq:mb_vih}
\end{equation}
where
\begin{equation}
	\frac{{\rm d}J_{\rm mb,Sk}}{{\rm d}t} = -3.8 \times 10^{-30}\;M_2\,R_\odot^4\,\left(\frac{R_2}{R_\odot}\right)^{\gamma_{\rm mb}}\omega_2^{3}\quad{\rm dyne\,cm}\;.
\end{equation}
We adopt $\gamma_{\rm mb} = 4$, $\tau_{\odot,{\rm conv}} = 2.8\times 10^{6}\;$s and $\dot{M}_{\odot,{\rm w}} = 2.54\times 10^{-14}\;M_{\odot}\,{\rm yr}^{-1}$. $\omega$ is the orbital angular velocity and $\tau_{\rm conv}$ is the turnover time of convective eddies. $\dot{M}_{\rm w}$ is the wind mass-loss rate of the secondary star. 

A caveat of the applied model MB3 is that \citet{cthc21} could not reproduce the many known wide-orbit binary MSPs using this magnetic braking prescription (because the drain of orbital angular momentum is simply too strong if applying MB3). Nevertheless, here we explore binary evolution using MB3 as our default model. 

We adopt the ``isotropic re-emission model'' for the mass transfer via RLO \citep[see e.g.][for a review]{tv06}.
In this model, the material flows conservatively from the donor star to the WD. The further fate of this material will strongly depend on the accretion rate \citep[see][]{cwyp+19}. If the accretion rate is larger than the maximum rate of steady burning, the excess material will be lost in the form of an optically thick wind \citep{hkn96}. Only in a narrow range of accretion, the H or He can burn stably, and the steady burning regimes of H and He do not overlap with each other \citep{nomo82b}. Therefore, even in the steady burning regime of H, not all material can be retained by the WD because of unsteady burning of He. If the accretion rate is lower than the minimum accretion rate of stable burning, the material will burning unstably and most of it will be ejected in violent shell flashes and nova outbursts and the WD may even reduce its mass if the accretion rate is too low \citep{hkn99}. Keeping this and also the low mass of the CO~WD of PTF~J0533+0209 in mind, we simply assume that, as an average for the entire RLO phase spanning a wide range of mass-transfer rates ($\dot{M}_2$), a fraction of $90\%$ of the transferred material is lost from the CO~WD accretor and takes away with it the specific orbital angular momentum of this WD.

The mass-transfer rate, $\dot{M}_2$ from the secondary star to the WD is computed with the scheme proposed by \citet{ritt88}. The amount of accretion onto the WD is limited by the Eddington accretion rate ($\dot{M}_{\rm Edd}$). We set the ratio of gravitational mass to rest mass of accreted material to be 0.99 for our CO~WD accretor. 

In our calculation, we also take the rotation of the donor star into account. We assume that its initial rotation velocity is synchronized with the initial binary orbital period. 
The prescription of \citet{pmsb+15} is adopted to compute the angular momentum loss due to spin-orbital couplings. The evolution of angular velocity is computed with the following formulae:
\begin{equation}
    \frac{{\rm d}\omega_{\rm j}}{{\rm d}t} = \frac{\omega_{\rm orb}-\omega_{\rm j}}{\tau_{\rm sync}}\;,
\end{equation}
and
\begin{equation}
    \frac{1}{\tau_{\rm sync}} = \frac{3}{q\,r_{\rm g}} \left(\frac{k}{T}\right)_{\rm c} \left(\frac{R}{a}\right)^{6}\;,
\end{equation}
where $\omega_{j}$ is the angular velocity at the face of cell $j$ and $\omega_{\rm orb}$ is the orbital angular velocity, $\tau_{\rm sync}$ is the synchronization time, $r_{\rm g}$ is the radius of gyration, $(k/T)_{\rm c}$ is the ratio of the apsidal motion constant to the viscous dissipation timescale, $R$ is the stellar radius and $a$ is the binary separation. The synchronization time depends on the structure of stars. For stars with convective envelopes, turbulent viscosity in the convective region, which may operate on the equilibrium tide, is the most efficient form of viscous dissipation. For stars with radiative envelopes, the resonances of the free gravity modes are damped by radiative dissipation which operates on the dynamical tide.
The value of $(k/T)_{\rm c}$ is different for these two kinds of stars and can be calculated as in \citet{htp02}. In our calculation, we do not include the internal heating due to the tidal fraction in the thermal evolution of the donors.

Chemical mixing and transport of angular momentum induced by different kinds of instabilities resulting from rotation \citep{hlw00} are also taken into account, i.e.: chemical mixing induced by secular shear instability, Eddington-sweet circulation, dynamical shear instability
and Goldreich-Schubert-Fricke instability. Following \citet{hlw00}, the mixing efficiency factor in our calculation is assumed to be $f_{\rm c} = 1/30$. With respect to the transport of angular momentum, we only consider the Spruit-Tayler dynamo \citep{spru02,hws05}. Following \citet{hlw00}, we set the parameter $f_{\mu}=0.05$, which regulates the inhibiting effect of chemical gradients on the mixing process. Motivated by \citet{imtl+16}, we also include element diffusion induced by gravitational settling and chemical and thermal diffusion in our calculation. 

\subsection{Initial values}
In order to find a model which can reproduce the properties of PTF~J0533+0209, we compute a grid of binary evolution with the above assumptions. Given the low mass of the currently observed CO~WD ($M_{\rm CO}=0.652\;M_\odot$), we assume an initial CO~WD mass of $M_{\rm CO,i}=0.55\;M_{\odot}$. The donor (secondary star) mass ranges between $M_{\rm 2,i}=1.0-1.4\;M_{\odot}$ in steps of $\Delta M_2 = 0.05\;M_{\odot}$. 
The initial orbital period ranges between $P_{\rm orb,i}=1.0-100\;{\rm d}$ in steps of $\Delta{{\rm log}(P_{\rm orb,i}/{\rm days})} = 0.1$. With this grid, we evolve models resulting in final properties relatively close to the properties of PTF~J0533+0209. Next, we decrease the steps of $\Delta P_{\rm orb,i}$ and find improved models which can better reproduce the observed properties of PTF~J0533+0209. Based on these models, we also compute some models with different WD or donor masses in order to understand the influence of these parameters on our results.

\subsection{Results}
\label{sec:res}
In this section, we present our binary evolutionary model with properties closest to those of PTF~J0533+0209. In this model, the initial values are $M_{\rm 2,i}=1.40\;M_\odot$, $P_{\rm orb,i}=7.00\;{\rm d}$ and $M_{\rm CO,i}=0.55\;M_\odot$.

\begin{figure}
    \centering
    \includegraphics[width=\columnwidth]{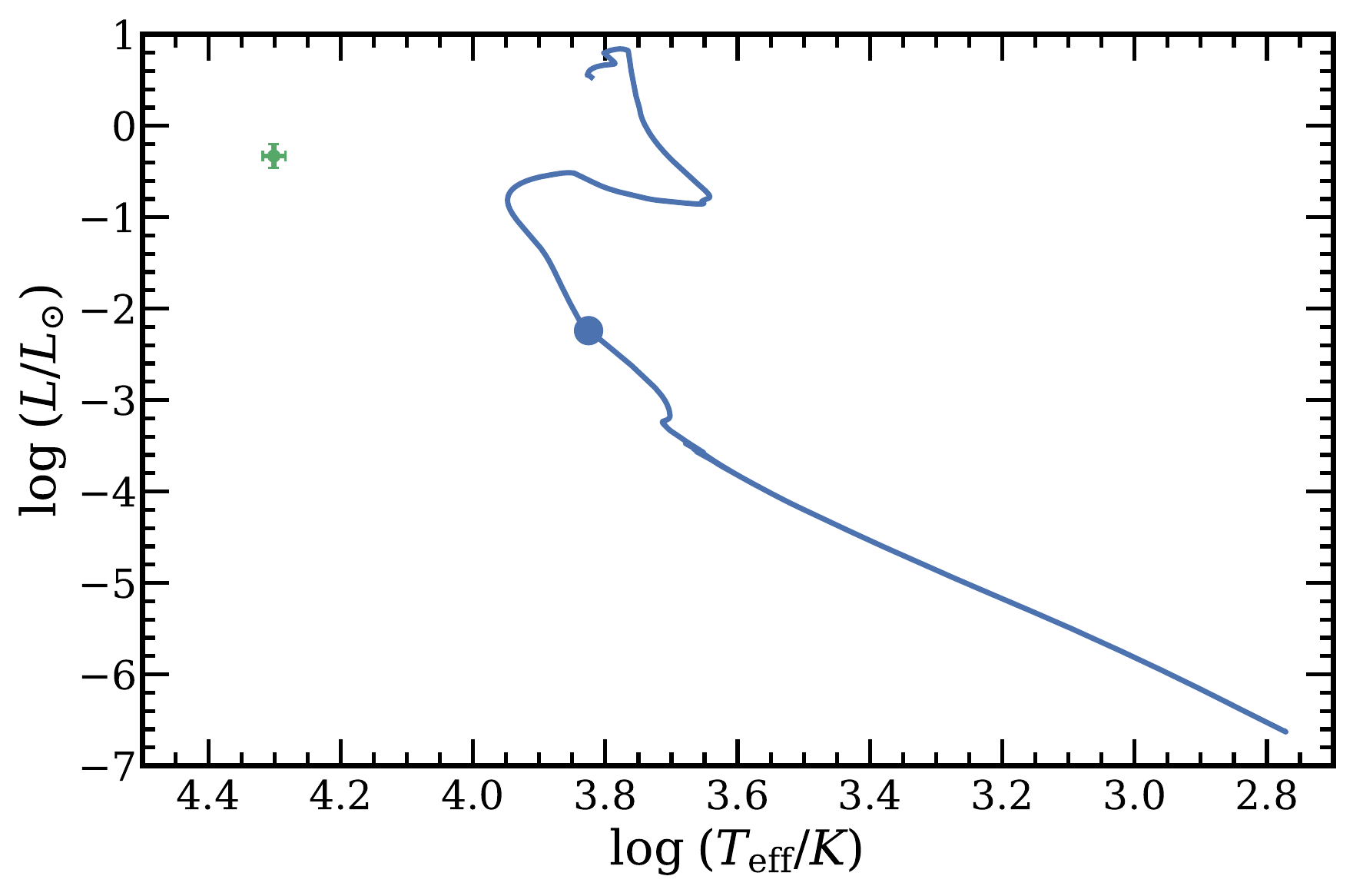}
    \includegraphics[width=\columnwidth]{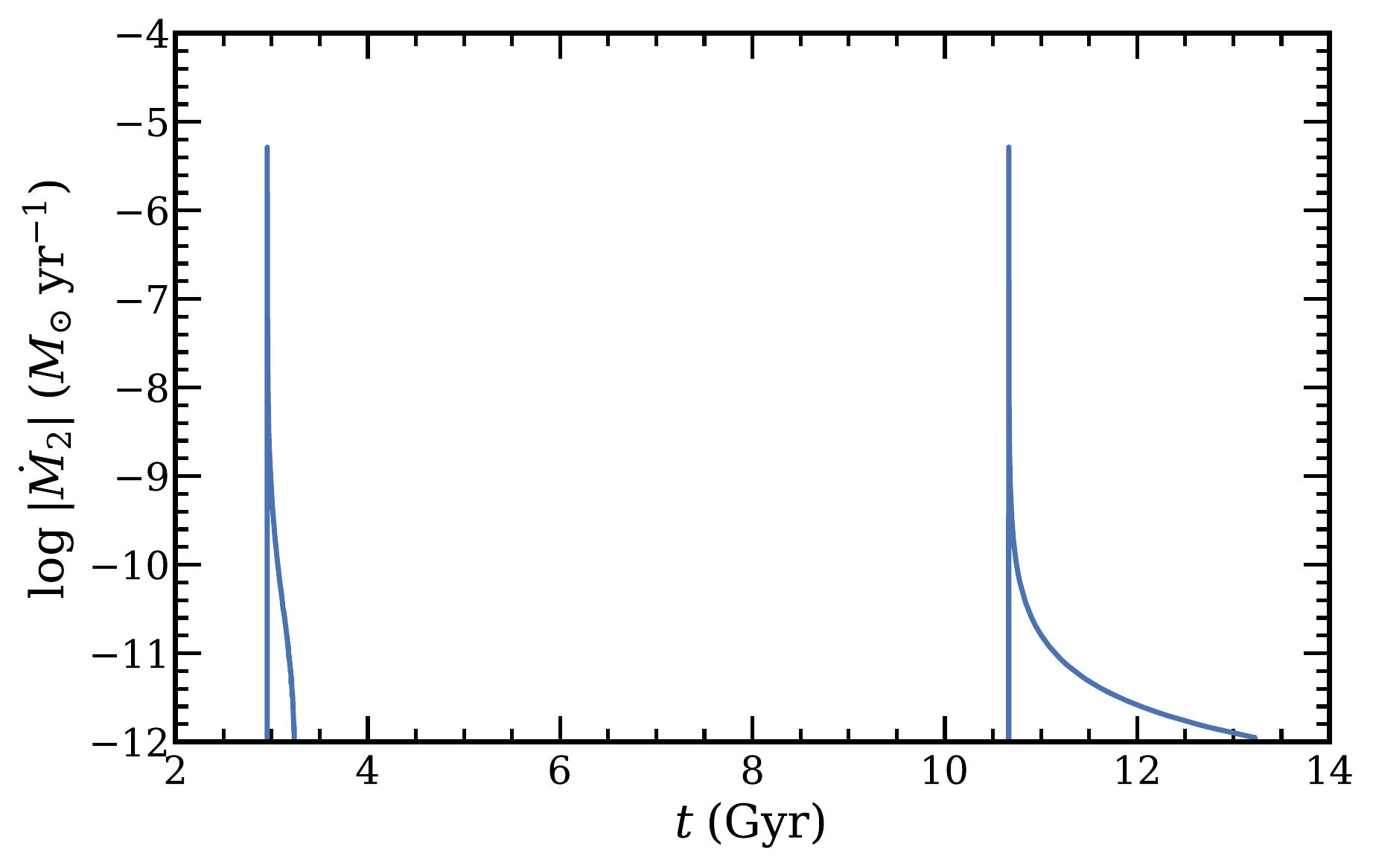}
    \includegraphics[width=\columnwidth]{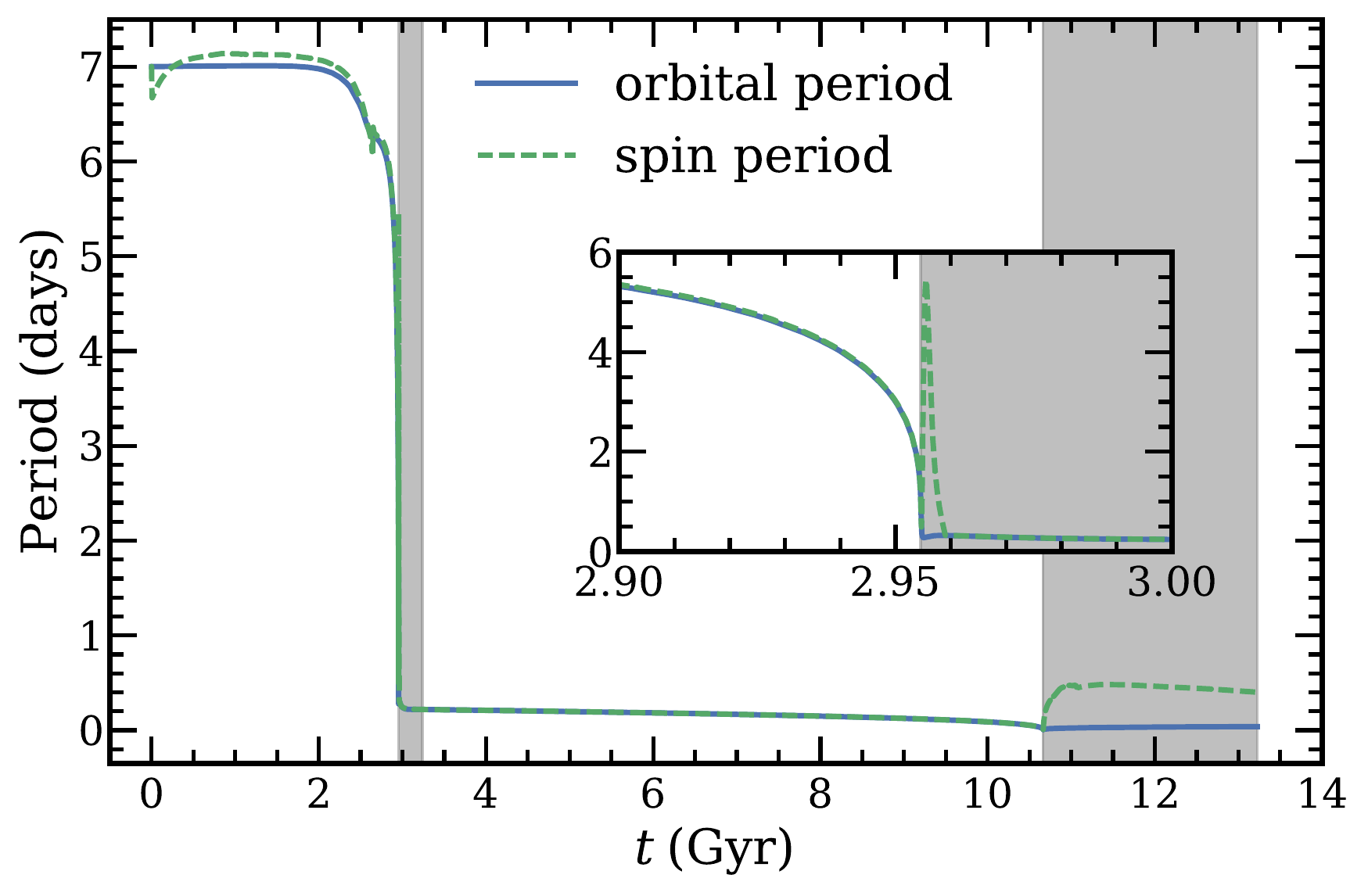}
    \caption{HR~diagram (top), mass-transfer rate (middle), orbital and spin period (bottom) of our best model for PTF~J0533+0209. The binary system was evolved with initial parameters $M_{\rm 2,i}=1.40\;M_\odot$, $P_{\rm orb,i}=7.00\;{\rm d}$ and $M_{\rm CO,i}=0.55\;M_\odot$. The blue dot in the plot represents the model which can well match the observed masses and orbital period of PTF~J0533+0209 (see Table~\ref{tab:para_com}), while the green dot with error bars indicates the actual location of PTF~J0533+0209. In the middle panel two phases of RLO is shown: CV and AM~CVn. Our evolved system has an age of {$t=10.660\;{\rm Gyr}$}. The middle panel shows the evolution of mass-transfer rate as a function of time. In the lower panel, the evolution of orbital (blue solid line) and spin period (green dashed line) are shown as a function of time. The two grey shaded regions indicate the two mass-transfer phases. The inset panel shows the evolution of spin and orbital periods around the onset of the first mass-transfer phase.}
    \label{fig:HR_mdot_ev}
\end{figure}

Figure~\ref{fig:HR_mdot_ev} displays the HR~diagram of the secondary star (top), the evolution of mass-transfer rate (middle), and orbital and spin periods (bottom), as a function of time. In the middle panel, two phases of RLO mass transfer are identified: cataclysmic variable (CV) and AM Canum Venaticorum (AM~CVn). PTF~J0533+0209 is located between these two stages.

A more qualitative description of this example is as follows: when the secondary star evolves beyond the main sequence (after termination of core hydrogen burning) it enters the Hertzsprung gap while its radius increases. Meanwhile (starting already from the ZAMS) the binary separation continuously decreases because of the magnetic braking. Therefore, the secondary star fills its Roche lobe and initiates mass transfer (and the system becomes a CV system) already at an age of $t = 2.95\;{\rm Gyr}$ when its radius is $R_2 = 2.53\;R_\odot$. 
This result can be understood as follows. Since the He~WD mass in PTF~J0533+0209 is quite small, its progenitor star should start mass transfer rather early in its stellar evolution shortly after it has produced a small core mass. And the ZAMS mass of the He~WD progenitor should not be too large, since in that case the RLO is no longer stable. In addition, early on the main sequence, the donor star is slightly out of synchronization. As the binary orbital period decreases significantly due to magnetic braking, the effect of tides becomes stronger and the spin of the donor star becomes synchronized with the orbit. At the very beginning of the first mass-transfer phase, because of mass loss, the donor star spins down and is out of synchronization with the orbit (see inset in bottom panel). However, due to the strong tides, it becomes synchronized again after a short time (less than 5~Myr).

At the end of the first mass-transfer phase ($t = 3.23\;{\rm Gyr}$), the secondary star detaches from its Roche lobe and slowly settles on the He~WD cooling track with a mass of $M_2 = 0.155\;M_{\odot}$. During the RLO, the CO~WD has accreted $\sim 0.12\;M_\odot$ and now has a mass of  $M_{\rm CO}=0.669\;M_\odot$ and the orbital period is $P_{\rm orb}=0.22\;{\rm d}$. 
The system subsequently evolves as a detached DWD system, as presently observed, with decreasing orbital separation due to GW radiation.
Around $t = 10.661\;$Gyr, the He~WD fills its Roche lobe and starts to transfer material onto the CO~WD and the system evolves as an AM~CVn system. Slightly before that stage ($t = 10.660\;$Gyr), when the orbital period and masses are close to those observed for PTF~J0533+0209, the system has the physical parameters listed in Table~\ref{tab:para_com}.

During the AM~CVn phase, the spin of the donor star is always out of synchronization with the orbit. This is because that the donor star spin increases due to the mass loss and the effect of tides becomes weaker as the orbital period increases.

\begin{figure*}
    \centering
    \includegraphics[width=\columnwidth]{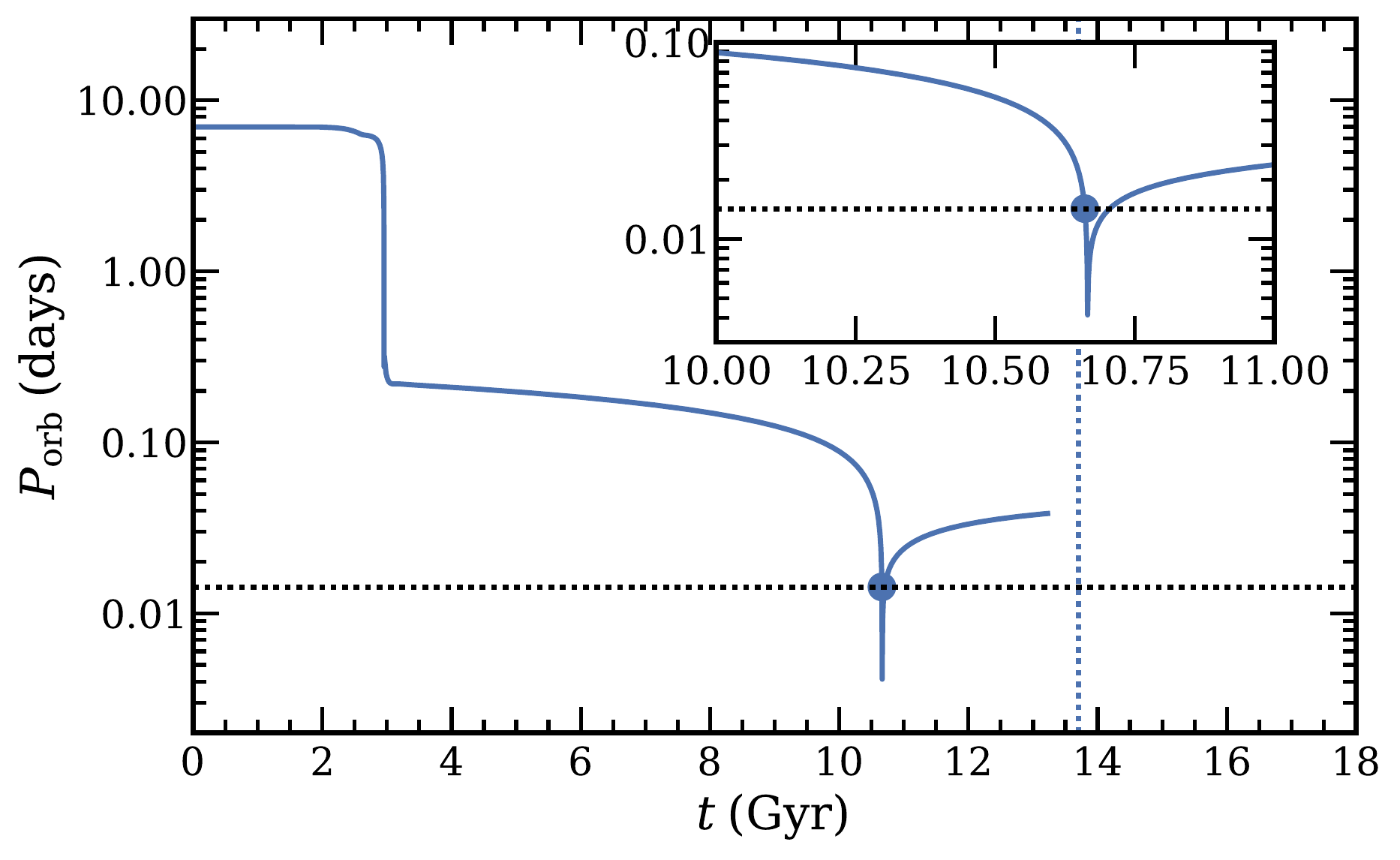}
    \includegraphics[width=\columnwidth]{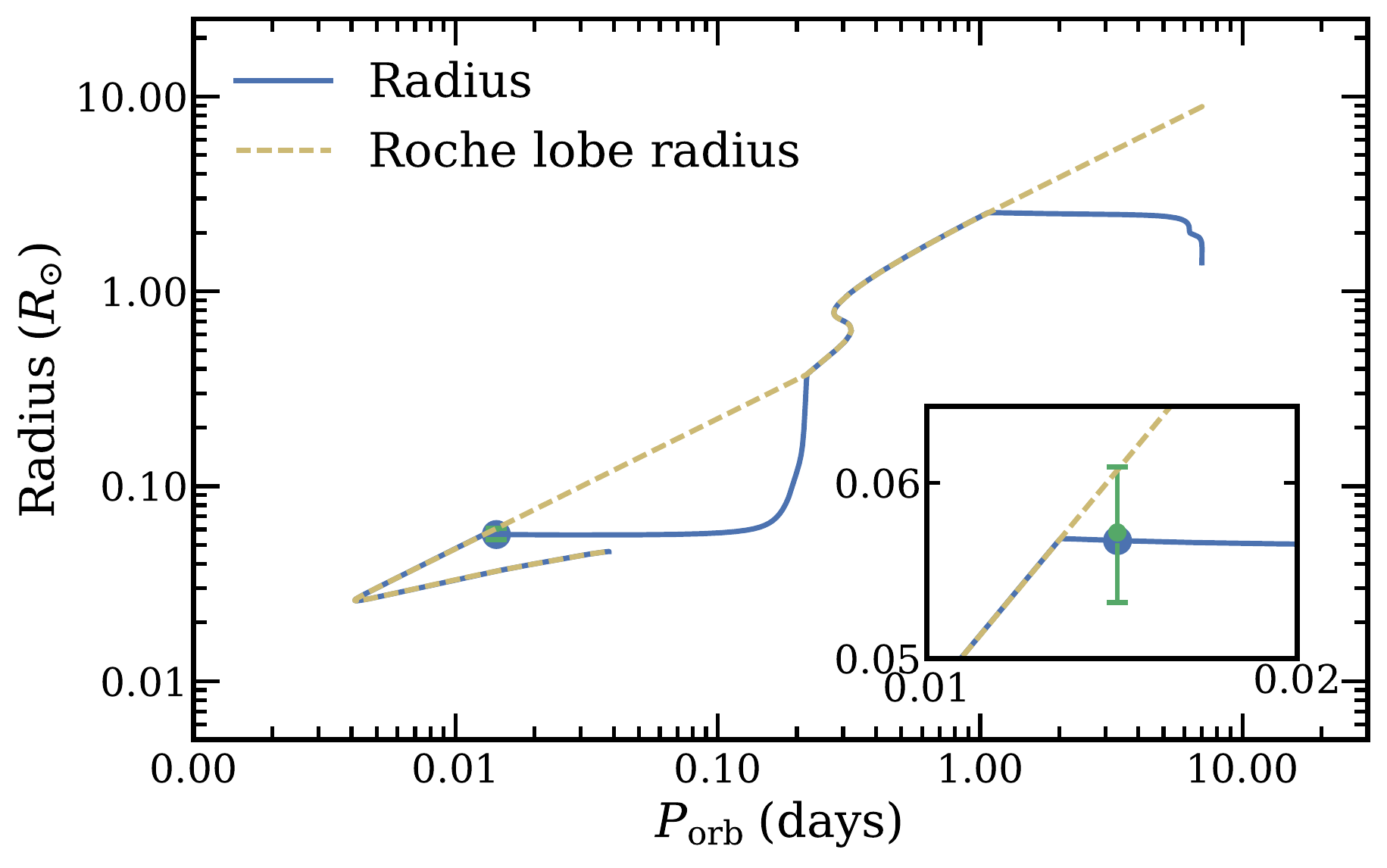}
    \includegraphics[width=\columnwidth]{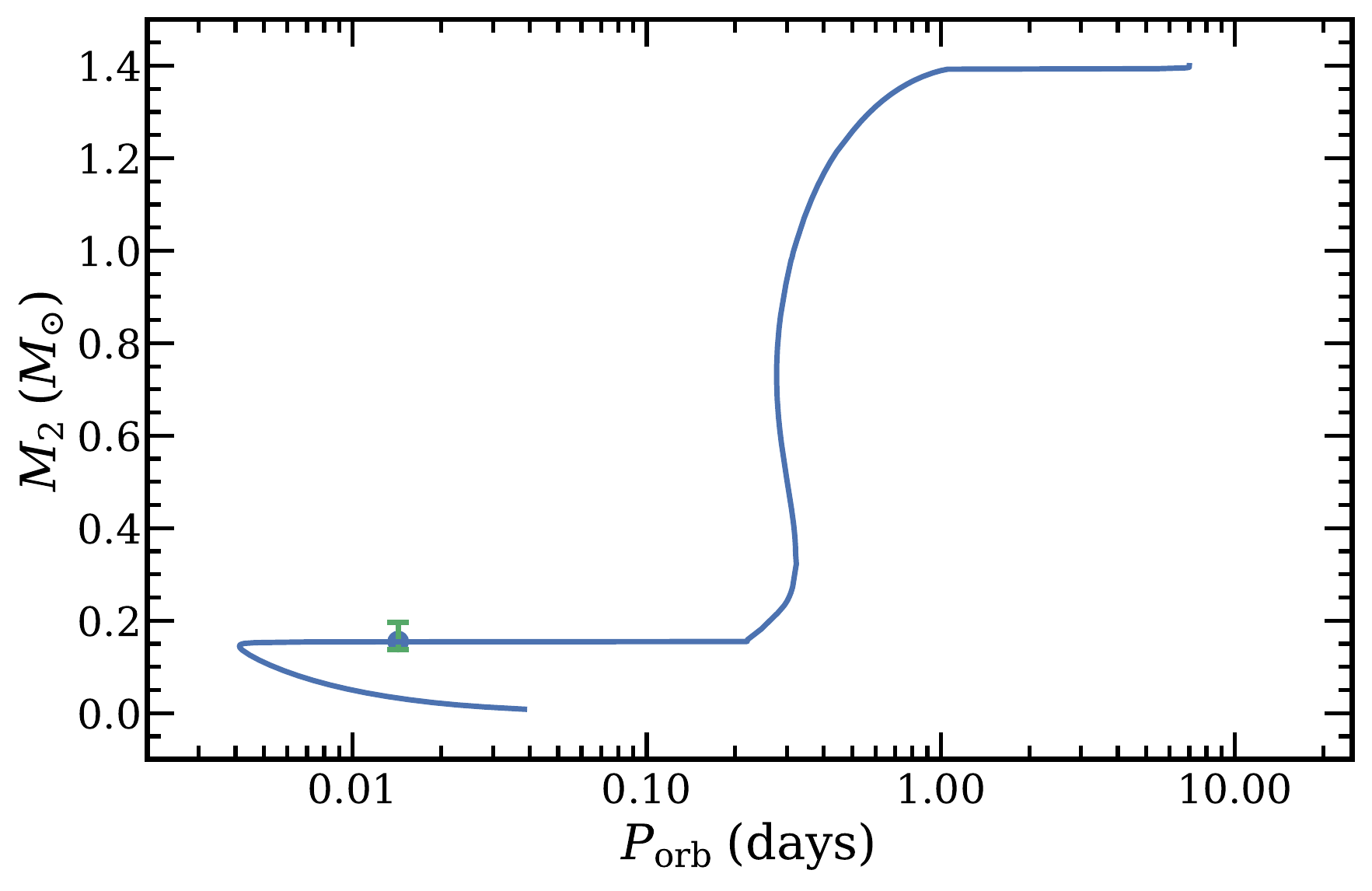}
    \includegraphics[width=\columnwidth]{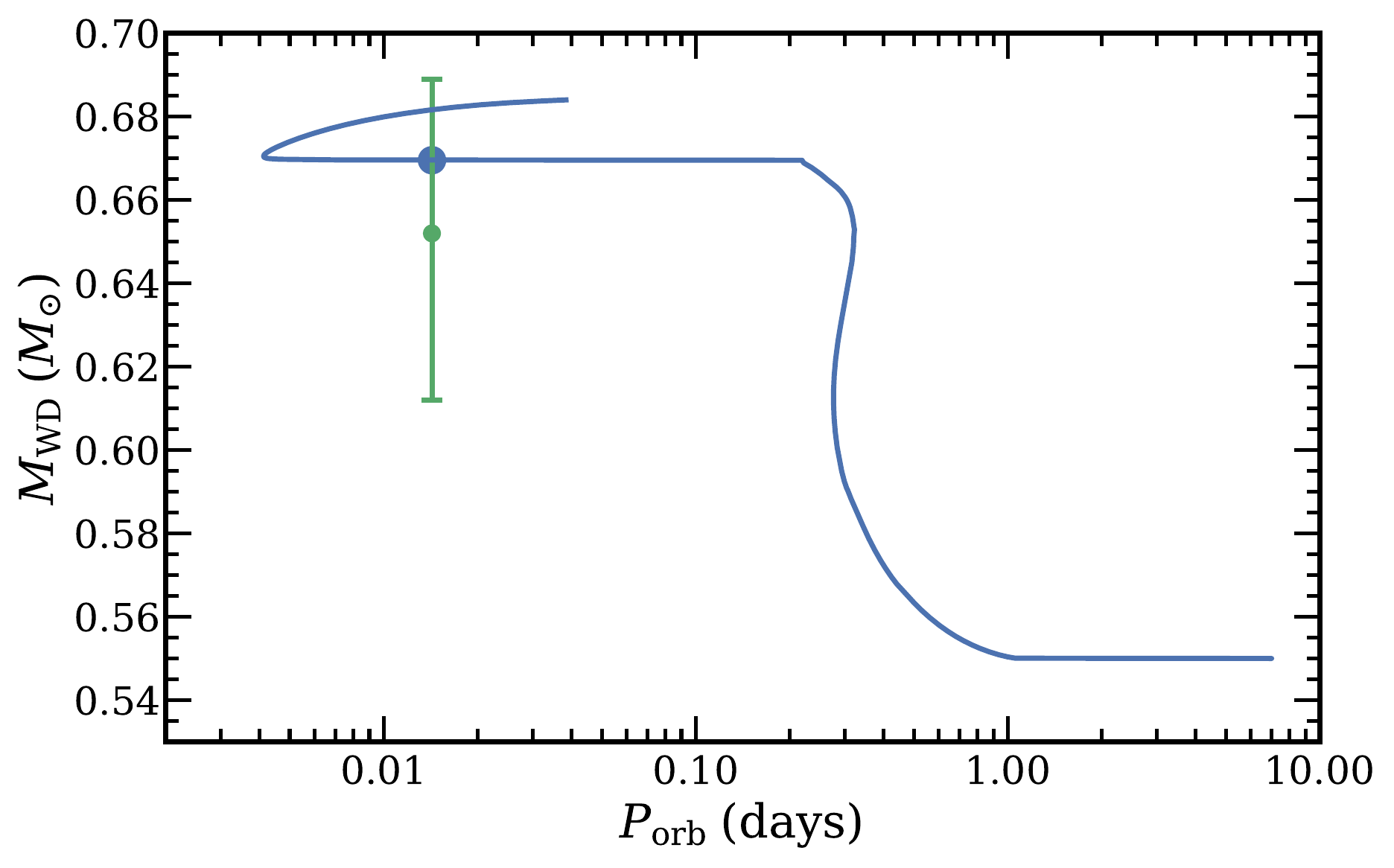}
    \includegraphics[width=\columnwidth]{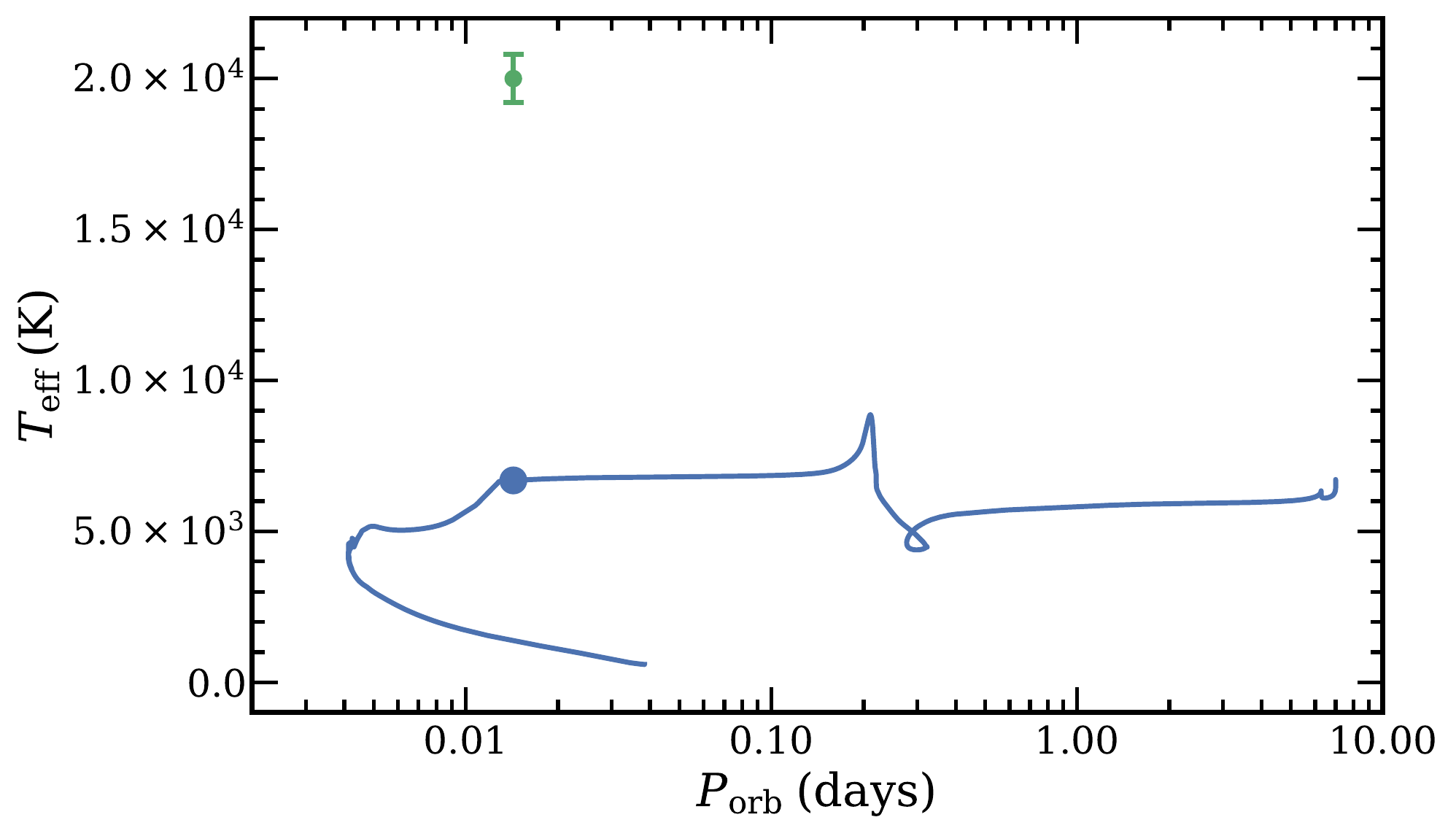}
    \includegraphics[width=\columnwidth]{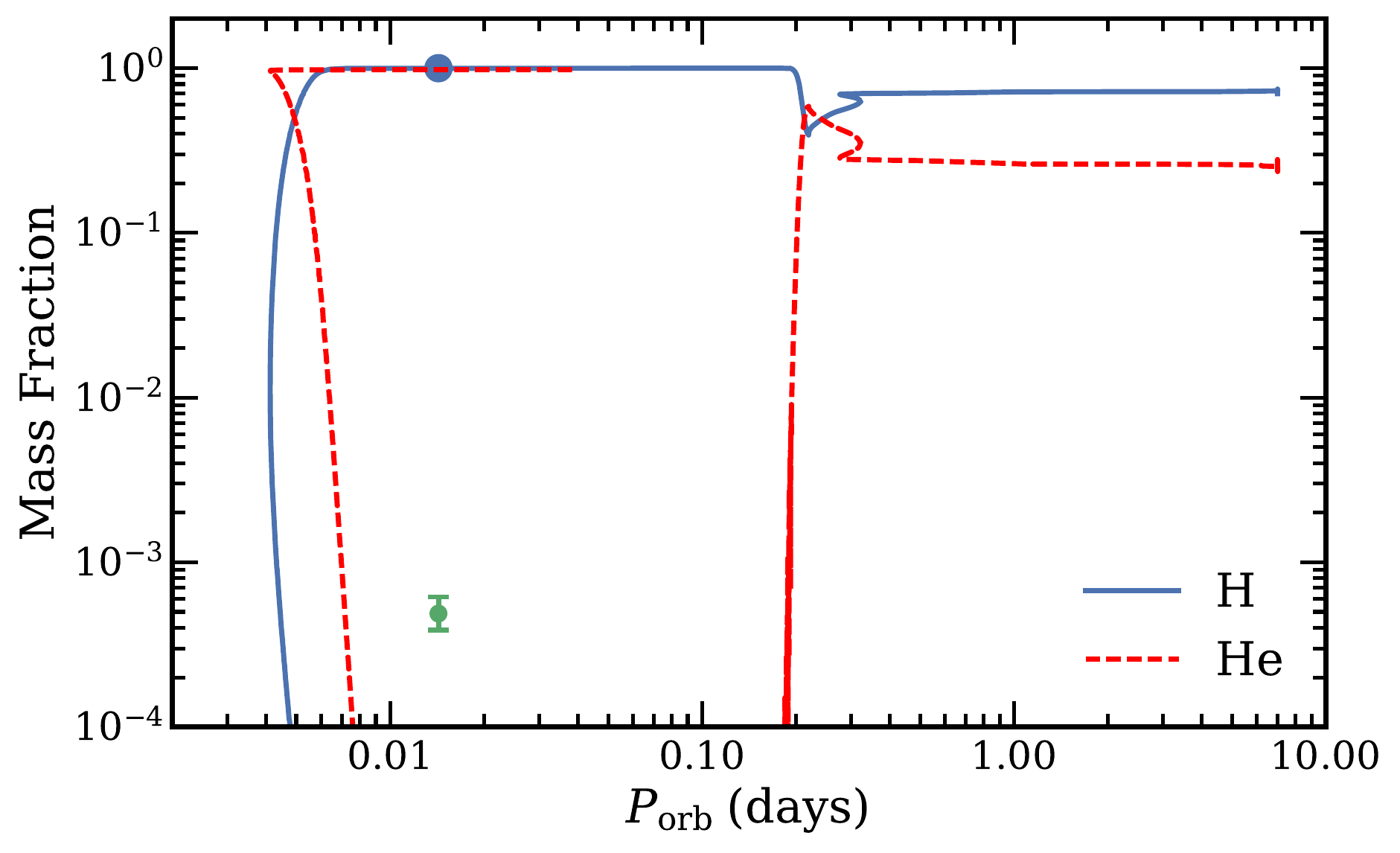}
    \caption{Comparison of our simulated system (best model values plotted as a blue circle) with the observed properties of PTF~J0533+0209 (plotted as a green dot with error bars, see Table~\ref{tab:para_com}). Upper-left panel: orbital period vs. stellar age. The vertical line indicates the Hubble time and the horizontal line indicates the observed orbital period of PTF~J0533+0209. Upper-right panel: donor star radius vs. orbital period. The blue solid and the yellow dashed lines show the evolution of stellar radius and Roche lobe size, respectively. The inset shows a zoom-in near PTF~J0533+0209. Central-left panel: donor star mass vs. orbital period. Central-right panel: mass of accreting CO~WD vs. orbital period. Lower-left panel: effective temperature vs. orbital period. Lower-right panel: donor star surface abundance vs. orbital period.
 }
    \label{fig:bin_evo}
\end{figure*}

Figure~\ref{fig:bin_evo} shows a more detailed comparison of our results with observations. Concluding from the values in Table~\ref{tab:para_com} and Fig.~\ref{fig:bin_evo}, we find that our model can very well reproduce the WD~masses and the orbital period of PTF~J0533+0209, as well as the ELM~He~WD radius (and $\log g$). However, the simulated effective temperature ($T_{\rm eff}$, and thereby also the luminosity, $L$) and the surface H-abundance ($X_{\rm surf}$) of the ELM~He~WD represent quite a mismatch. In the grid we computed, we have checked if other models can better reproduce $T_{\rm eff}$ and $X_{\rm surf}$. However, we conclude that this is not the case (see discussions in Section~\ref{sec:dis}).

For our initial binary systems, if we increase the donor star mass ($M_{\rm 2,i}$) or the orbital period ($P_{\rm orb,i}$), we produce 
He~WDs with too large masses and too small radii.
If we decrease $M_{\rm 2,i}$ or $P_{\rm orb,i}$, the results is that the timescale of nuclear evolution is too long to reproduce PTF~J0533+0209 or the system never detaches from RLO \citep[producing a so-called converging system, see e.g.][]{itl14}, respectively. 
If we increase the initial CO~WD mass ($M_{\rm CO,i}$) of the accretor, the results are not very significantly different. To compensate for a different applied value of $M_{\rm CO,i}$ we can simply adjust the accretion efficiency to reach the value of $M_{\rm CO}=0.652\;M_\odot$ for PTF~J0533+0209). 

\section{Discussion} 
\label{sec:dis}

\subsection{A possible way to mitigate the discrepancy}
As we found in the previous section, our model cannot explain the high value of $T_{\rm eff}$ and the low H~abundance of the ELM~He~WD in PTF~J0533+0209. In the following, we will discuss the possible explanation for this discrepancy. However, we start by briefly summarizing the investigation by \citet{bfpv+19}.

The peculiar observed values of the above-mentioned parameters were also discussed by \citet{bfpv+19} in their attempt to model PFT~J0533+0209. Whereas we have argued that a CE origin of PTF~J0533+0209 seems unlikely based on simple energy considerations (Section~\ref{sec:ce}), they disregarded this aspect and instead attempted to explain PTF~J0533+0209 by modelling a post-CE system in composite steps: first, they evolved a single $1.20\;M_\odot$ star up the red-giant branch until its core mass is $0.187\;M_\odot$. They then stripped the outer layers until the star had a mass of $0.19\;M_\odot$ and a total amount of hydrogen of $10^{-3}\;M_\odot$. Subsequently, they placed this model star in a $P_{\rm orb}=1\;{\rm hr}$ binary system with a $0.66\;M_\odot$ companion and evolved it further using \texttt{MESA}. Despite introducing excessive rotational mixing, the surface abundance of hydrogen remained too high.
They then computed a model with an extremely large ad-hoc mixing diffusivity, which enforces hydrogen to mix deep enough to burn and thereby reduce the value of $X_{\rm surf}$. 

\citet{bfpv+19} concluded that based on the findings of \citet{fl13}, the tidal heating is unlikely to heat up the ELM~He~WD to the observed temperature at an orbital period of 20~min. Referring to the work of \citet{bcf+19}, they argued that the ``tidal temperature'' to which the ELM~He~WD can be heated is only $\sim 7000\;{\rm K}$, much below the observed temperature of $T_{\rm eff}\sim 20\,000\;{\rm K}$.
Another solution discussed by \citet{bfpv+19} is the possibility of H-shell flashes \citep[e.g.][]{dsbh98,asb01,ndm04,amc13,imtl+16}. 
However, due to numerical reasons, they were not able to simulate flash-induced RLO of the residual H-envelope. Such RLO could substantially reduce the amount of residual hydrogen (see below).
Finally, \citet{bfpv+19} attempted to model PTF~J0533+0209 via the stable RLO channel as explored in this paper. However, due to the problems of reproducing $T_{\rm eff}$ and $X_{\rm surf}$, they concluded that the CE channel is more likely.

Despite the initial issues with $T_{\rm eff}$ and $X_{\rm surf}$, we are still optimistic with respect to the stable RLO channel. 
The residual H-envelope mass expected for ELM~He~WDs is typically between $\sim 10^{-3}-10^{-2}\;M_\odot$ depending on WD mass and evolutionary epoch. 

\begin{figure*}
\centering
\includegraphics{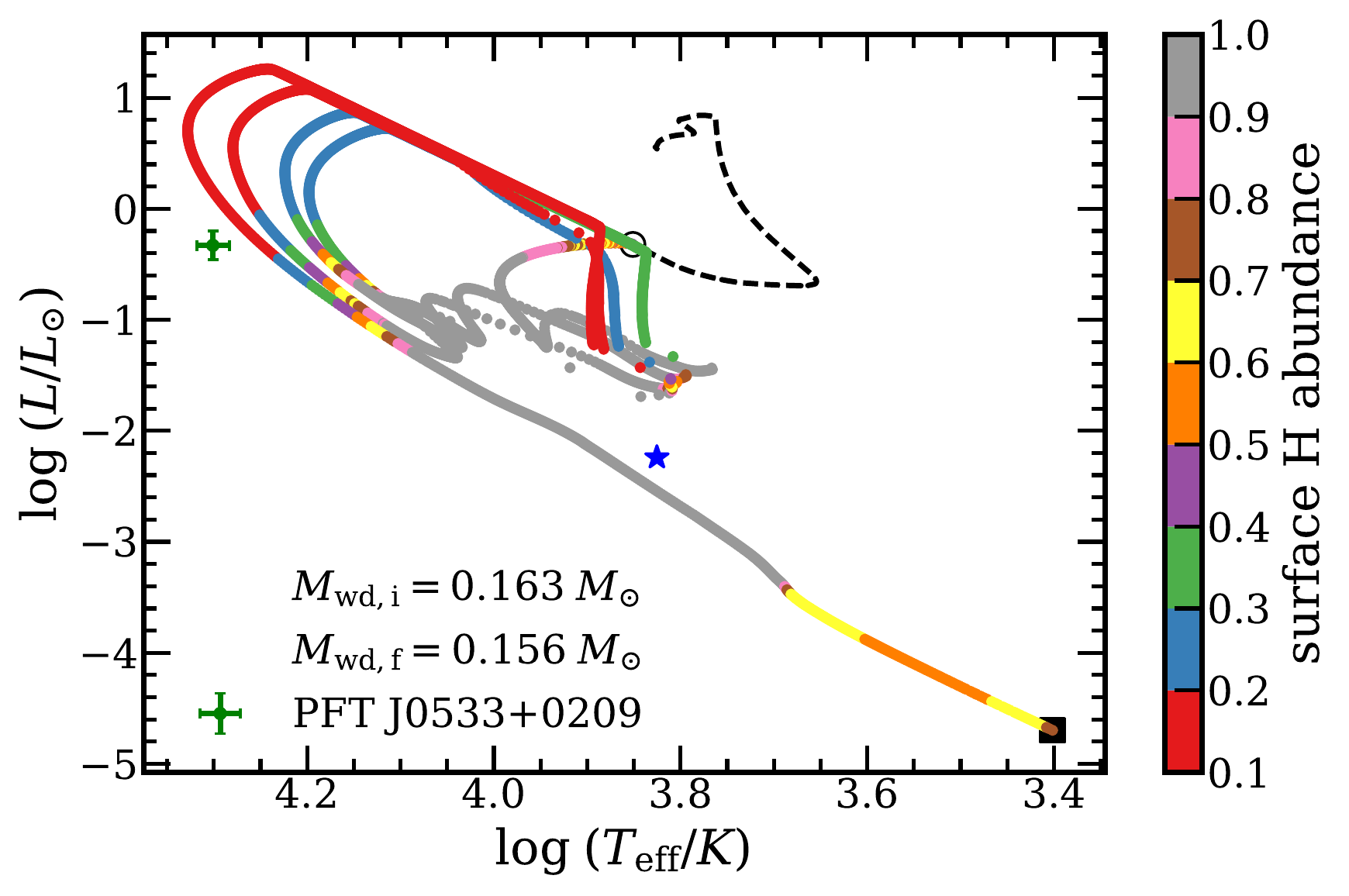}
\caption{Evolutionary model of a (final) $0.156\;M_\odot$ ELM~He~WD from our grid. In this model, the initial binary parameters are $M_{2,i} = 1.40\;M_{\odot}$, $M_{\rm CO, i} = 0.63\;M_{\odot}$, $P_{\rm orb, i} = 8.6\;$d and accretion efficiency $0.0$. 
Shown here are the \texttt{MESA} models which cover the evolution from detachment of the CV~phase ($t=3.2\;{\rm Gyr}$, shown with a black open circle) until $t=13.7\;{\rm Gyr}$ and the end of the WD cooling track (black square).
The black dashed line shows the evolution of the progenitor of the ELM He WD from the ZAMS until the end of CV phase.
Several large H-shell flashes (loops) are seen, including some with RLO. The color on the evolutionary track represents the surface H abundance (see color bar). 
The green dot with the error bar represents the current location of PTF~J0533+0209. Our model presented in Table~\ref{tab:para_com} is shown with a blue star.
}
\label{fig:flashes}
\end{figure*}

Figure~\ref{fig:flashes} displays the HR~diagram of one of our grid model computations of the evolution of an ELM~He~WD model with a final WD mass of $0.156\;M_\odot$. It covers the evolution from detachment of the CV~phase as a $0.163\;M_\odot$ proto-ELM~He~WD ($t=3.2\;{\rm Gyr}$ and $P_{\rm orb}=6.92\;{\rm hr}$) until $t=13.7\;{\rm Gyr}$ ($P_{\rm orb}=5.97\;{\rm hr}$). 
The WD experiences several strong H-shell flashes during its evolution, leading to temporary episodes of additional RLO. As a consequence, the final WD mass is reduced to $0.156\;M_{\odot}$. From the plot it is seen that following each flash episode the surface H-abundance is considerably reduced, although in all cases $X_{\rm surf}\ga 0.10$. Thus, we cannot explain the observed value of $X_{\rm surf}\simeq 4.9\times 10^{-4}$, which may indeed suggest that extra mixing is needed, as proposed by \citet{bfpv+19}. Nevertheless, we speculate, that it is quite possible that PFT~J0533+0209 is currently observed shortly after one of these short-lived H-shell flash episodes, thereby helping to explain the measured values of $T_{\rm eff}$ and $L$. A caveat is that both $X_{\rm surf}$ and $L$ ($T_{\rm eff}$) decay back to pre-flash values on a timescale of order 10~Myr. Hence, the time window for observing the system in an immediate post-flash evolutionary stage is rather limited.

\begin{figure}
\centering
\includegraphics[width=\columnwidth]{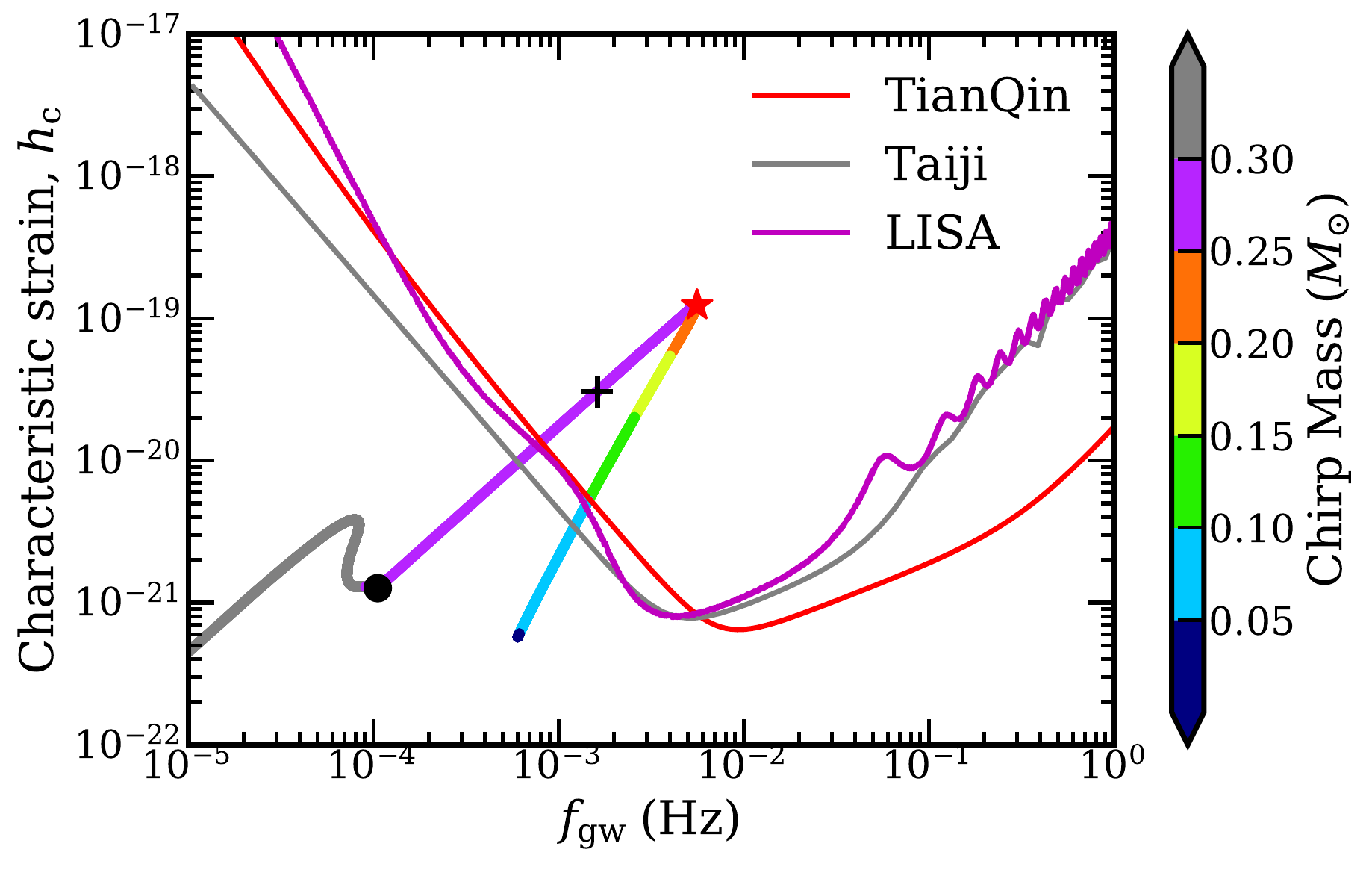}
\caption{Evolution of the characteristic strain as a function of GW frequency for the binary system presented in Fig.~\ref{fig:HR_mdot_ev} for an assumed distance of $d=1.5\;{\rm kpc}$.
The red, grey and magenta solid lines represent the sensitive curves for TianQin with an observational time $T = 5\;$yr, Taiji with an observational time $T = 5\;$yr and LISA with an observational time $T = 4\;$yr, respectively. The color on the evolutionary track represents the chirp  mass of the binary system (see color bar; the grey and dark blue colors represents chirp masses of $M_{\rm chirp} > 0.30\;M_{\odot}$ and $M_{\rm chirp} < 0.05\;M_{\odot}$, respectively).
The black dot represents the end of first mass transfer phase and the star represents the onset of the second mass transfer phase. The cross represents the current location of PTF~J0533+0209.
}
\label{fig:fre_str}
\end{figure}

\subsection{Influence of the accretion disk}
In our calculations, we did not consider the effect of an accretion disk during the AM~CVn~phase. In this phase, the binary separation is very small and the formation of accretion disk is not obvious. In an early paper, \citet{npvy01} studied this question with an analytical method. If the minimum distance at which the accretion flow passes the accretor is comparable to the radius of the accretor, it is likely that no accretion disk will be formed and the accretion flow will directly hit the surface of the accretor. 
From their fig.~1, it is suggestive that no accretion disk is formed at the beginning of the AM~CVn phase whereas an accretion disk will appear at later times, after the system passes through the orbital period minimum and once the donor star (He~WD) mass is below $\sim 0.08\;M_\odot$.
Without an accretion disk, loss of angular momentum from the orbit may destabilize the mass transfer unless the angular momentum lost to the accretor can be transferred back to the orbit \citep{mns04}. 
Later, it was argued by \citet{kbs12} that for AM~CVn systems with ELM He~WD donors, the initial contraction of the ELM~He~WD due to mass loss allows for more stable mass transfer than that originally found for cold He~WDs studied by \citet{mns04}.
In our \texttt{MESA} models, we do take into account the entropy (finite temperature) of the ELM~He~WD, but the detailed treatment of direct impact versus disk formation is not included.

\subsection{A LISA verification source}
PTF~J0533+0209 is expected to be detected by LISA \citep[i.e. a so-called LISA verification source][]{kks+18}. However, during the subsequent GW damping of the system, the more extended ELM~He~WD component will fill its Roche lobe and initiate RLO towards the CO~WD companion. During this epoch, the system will be observable as an AM~CVn system \citep{npvy01,vkg+21}. Discussions and detailed tracks of their evolution through the LISA GW frequency band during this phase is given in e.g. \citet{taur18}.  

Figure~\ref{fig:fre_str} shows the evolution of characteristic strain as a function of GW frequency for our binary model shown in Fig.~\ref{fig:HR_mdot_ev}. In the calculation, we assume the distance of the system to be $d = 1.5\;$kpc following \citet{bfpv+19}.
Besides LISA, there are other low-frequency GW observatories which will be lunched in the next two decades, such as TianQin \citep{lcdg+16} and Taiji \citep{rgcz20}. In the plot, we therefore also present the sensitive curves for the GW observatories TianQin, assuming an observational time $T = 5\;$yr, Taiji with an observational time $T = 5\;$yr, besides LISA with an assumed observational time $T = 4\;$yr. 
We notice that a system like PTF~J0533+0209 is detected both during the detached phase and the subsequent AM~CVn phase. 
From this plot, we find a possible signal-to-noise ratio (S/N) up to $10-100$ for simulated systems similar to, but further evolved than, PFT~J0533+0209. The chirp mass of the detectable sources are between $0.10-0.27\;M_{\odot}$ (steadily decreasing with evolutionary age, see right-hand side colour bar).

\section{Conclusions}
\label{sec:con}
In this paper, we have discussed possible formation scenarios of the DWD binary system PTF~J0533+0209. We found that explaining its origin via the CE channel \citep[as suggested by][]{bfpv+19} is difficult since the low mass of the ELM~He~WD implies CE ejection at an early evolutionary stage (Hertzsprung gap) when the binding energy of the envelope of the progenitor star is still larger (by a factor of $\sim 2$) than the amount of orbital energy released during spiral-in. 

We have instead presented a formation scenario based on stable RLO in a CV system. Using the \texttt{MESA} code in combination with a magnetic braking prescription from \citet{vih19}, we have modelled the system from the ZAMS and found a solution that matches the two measured WD component masses and the present orbital period of the system. However, our model fails to reproduce the observed $T_{\rm eff}$ and, in particular, the surface H-abundance of the ELM~He~WD. We have discussed possible reasons for this and point to a recent H-shell flash (combined with efficient mixing) as the most plausible explanation. 

The present GW signal of this system will be detected by low-frequency GW observatories such as LISA, TianQin and Taiji. We calculated the signal of its detection and continued our computation for PTF~J0533+0209 during its future AM~CVn phase when the ELM~He~WD fills its Roche lobe in a bit more than $1\;$Myr from now.

We are strongly inclined to believe that PFT~J0533+0209 formed via stable RLO. However, a complete and self-consistent model that explains all observables of the binary is still missing. We consider this system a very interesting test case for further investigations of close binary star evolution and input physics, and encourage the community to join our efforts.

\begin{acknowledgments}
We thank the anonymous reviewer for constructive comments.
H.-L.C. thanks Ren Song and Zhenwei Li for helpful discussions. 
H.-L.C. gratefully acknowledges support and hospitality from Aarhus University. This work is partially 
supported by the National Natural Science Foundation of China (Grant No. 12090040, 12090043, 11521303, 12073071, 11873016, 11733008), 
Yunnan Fundamental Research Projects (Grant No. 202001AT070058, 202101AW070003), the science research grants from the China Manned Space Project with No. CMS-CSST-2021-A10) and Youth Innovation Promotion Association of Chinese Academy of Sciences (Grant no. 2018076). H.-L.C. acknowledges the computing time granted by the Yunnan Observatories
and provided on the facilities at the Yunnan Observatories Supercomputing Platform. 
We are grateful to the \textsc{mesa} council for the \textsc{mesa} instrument papers and website. 
\end{acknowledgments}

\vspace{5mm}
\software{MESA \citep{pbdh+11,pcab+13,pmsb+15,psbb+18,pssg+19}  
          }

\bibliography{ms.bbl}{}
\bibliographystyle{aasjournal}

\end{document}